# A Hybrid Approach for an Interpretable and Explainable Intrusion Detection System

Tiago Dias[0000-0002-1693-7872], Nuno Oliveira[0000-0002-5030-7751], Norberto Sousa[0000-0003-2919-4817], Isabel Praça[0000-0002-2519-9859], and Orlando Sousa[0000-0003-0779-3480]

Research Group on Intelligent Engineering and Computing for Advanced Innovation and Development (GECAD), Porto School of Engineering (ISEP), 4200-072 Porto, Portugal
{tiada,nunal,norbe,icp,oms}@isep.ipp.pt

**Abstract.** Cybersecurity has been a concern for quite a while now. In the latest years, cyberattacks have been increasing in size and complexity, fueled by significant advances in technology. Nowadays, there is an unavoidable necessity of protecting systems and data crucial for business continuity. Hence, many intrusion detection systems have been created in an attempt to mitigate these threats and contribute to a timelier detection. This work proposes an interpretable and explainable hybrid intrusion detection system, which makes use of artificial intelligence methods to achieve better and more long-lasting security. The system combines experts' written rules and dynamic knowledge continuously generated by a decision tree algorithm as new shreds of evidence emerge from network activity.

## 1 Introduction

With the latest advances in technology, lots of sensitive information from both people and enterprises are shared across a wide variety of devices through the internet. Computational technologies lack security, leaving attackers numerous opportunities to exploit systems that may contain sensitive data that can be crucial for the business continuity of an enterprise. According to a study made in May of 2019, "Cyber and Physical Security: Perspective from the C-Suite", CEOs and corporate boards are putting more effort to increase cybersecurity [1]. To increase security in networks, there are multiple protocols/frameworks that can be followed and tools that can be used. An enterprise may follow specific security frameworks, which will raise various use cases concerning security. Depending on the use cases, the network security administrator should be able to understand which tool suits best.

Intrusion Detection Systems (IDSs) are important security mechanisms that analyze the traffic that flows through networks and systems to identify and alert suspicious activity, resorting to anomaly detection methods [2]. According to S.KishorWagh et al. in [3], there are three types of anomaly detection. These are statistical, knowledge, or machine learning-based. The first one analyzes how the net-work usually behaves, and any perturbation of that state can be considered an anomaly. The knowledge-based approach, the most used one [3], takes domain information provided by a knowledge engineer, with the assistance of an expert, in the form of domain rules and populates a knowledge base that an inference engine can utilize to detect anomalies. A knowledge base is a store of facts, assumptions, and rules computed to solve domain problems. The most recent approach, machine learning-based anomaly detection, uses Machine Learning (ML) models to detect anomalies. These are typically supervised or unsupervised. Regarding the first, it usually goes through a training phase, in which it takes data to define the ML model's behaviour, being this an essential step of the model's performance. In contrast, the unsupervised models do not require training. Instead, they may resort to other methods such as clustering or rewarding agents for the sake of learning.

Nowadays, most IDSs, such as Snort [4], Suricata [5], and Zeek [6], are rule-based but lack adaptability, hence the recent research on using machine learning-based approaches to detect anomalies. Unfortunately, although the latter may have great results, they leave users ignorant regarding the detection process since they do not know why or how an attack was identified, as most of them rely on black box models.

This work aims towards enhancing the cybersecurity of network assets by proposing an IDS detent of ML-based features. The proposal follows a rule-based approach, meaning that it is as robust and resilient as the knowledge base is complete. However, it is hard to maintain and expand it with valuable rules, hence the introduction of the ML to this context. Furthermore, it raises the capability of deriving new knowledge from empirical evidence, which may be helpful to expand the knowledge base, and therefore helpful to the Security Operations Center (SOC) Engineer that manages the system's knowledge base.

This paper is organized as follows. Section 2 provides a literature review about existing IDSs, an overview of explainability and interpretability concepts and machine learning applications to IDSs. Section 3 analyses this work's contributions, an architectural and functional understanding of the proposed system. Section 4 presents a case study and the results achieved. Finally, section 5 summarises the work with the main conclusions drawn from it and describes new research lines to be explored in the future.

## 2      Related Work

Over the years, many IDSs have been developed to increase security over networks. Attempts to merge these existing IDSs with ML models have also been made. This chapter provides a literature review about existing IDSs and reviews artificial intelligence applications to the cybersecurity field.
IDSs are network security mechanisms that allow users to be aware of possible network intrusions by raising alerts. An IDS can be of multiple types, the most common being host-based IDS (HIDS) and network-based IDS (NIDS). Regarding the latter, there are some open-source IDSs such as Snort, Suricata, and Zeek.

These IDSs detent certain capabilities, but they are commonly working towards the same goal, intrusion detection. All of them use a knowledge-based approach for detecting anomalies, adopting anomaly-based and signature-based methods to perform intrusion detection. Particularly, Zeek goes beyond detection, as it includes performance measurement and troubleshooting. Also, its architecture defines a metamodeling concept that allows a knowledgeable individual to extend it by modelling business-specific scenarios.

Overall, Zeek ends up being the most versatile one since it can detect a broader range of activity patterns, but it is harder to configure. On the other hand, Snort and Suricata are more intuitive and thoroughly. The question of which or what kind of IDS should be deployed always depends on the size and scale of the organization's internal networks [2]. The advantage of using various IDSs simultaneously is that different IDSs, by default, may possess different capabilities for detecting anomalies. This can help security administrators determine missed vulnerabilities and exploits that could be performed by an attacker [2].

Considering the National Institute of Standards and Technology (NIST) [7] cybersecurity framework [8], the IDSs mentioned above fulfil the identification and detection pillars. However, they can be expanded and therefore comply with the protection, response and recovery guidelines.

As a result of these characteristics, research has been made to understand how machine learning techniques could improve these systems.

According to recent literature [9], the detection capabilities of AI algorithms studied, such as Random Forests (RF), Multi-Layer Perceptron (MLP), Long-Short Term Memory (LSTM), Decision Trees (DT), k-Nearest Neighbors (kNN) and Support Vector Machines (SVM), work considerably well. For instance, Oliveira et al. [9] conducted an experiment with RFs, MLPs, and LSTMs to understand which has the best performance by comparing single-flow and multi-flow detection approaches in the CIDDS-001 dataset. They concluded that anomaly-based intrusion detection for CIDDS-001 was better addressed from a multi-flow perspective, classifying the LSTM model as the most promising one.

However, the AI algorithms mentioned above are typically black-box, which can be an embargo on achieving explainability [10]. Interpretability and explainability are challenging assets to achieve in a system. It is significantly more complex when it uses ML, as most of them use black-box techniques. These are essential for users to effectively understand, trust and interact correctly with systems, even more if these are AI-based. Human-centric IDSs require human interaction with the agent. The reason for explainability, in this case, is to persuade the designated human participant to interact with it [11].

The same would apply if a knowledge-based IDS were to become a unified version of a human-centric and agent-centric system by containing knowledge discovering methods [11]. For an ML model to be fully understood, a human should be able to perform every calculation to produce a prediction [11][12]. Therefore, research has been made to test explainable models for detecting anomalies.

Mahbooba et al. [10] decided to take the DT approach to ensure interpretability since it provides a human-based approach for decision-making, concluding that the algorithm was able to rank the importance of features, provide explainable rules achieving an accuracy similar to state-of-the-art algorithms.

Some authors have decided to research the integration of the aforementioned NIDS with some of the above-mentioned ML models. S.Shah et al., in their article [13], decided to take Snort and Suricata and enhance them by testing their integration with ML. They performed several tests on the IDSs without the support of ML to understand their native flaws. They proceeded to test five different algorithms to find the most appropriate one and were able to conclude that an optimized version of the SVM, using the firefly algorithm, achieved the best results.

V. Gustavsson, in his dissertation [14], used Zeek to extract features and send them to ML models to be classified. He tested the tool with already existing detection scripts without ML. Then proceeded to test the IDS's speed with and without flow feature extracting scripts. Lastly, the author tested each ML model to understand which option was the best to integrate with Zeek. Based on the metrics extracted, KNN showed the best results.

C. Sinclair et al. [15] implemented a module that contained interfaces with a Genetic Algorithm (GA) and a DT. The authors used a DT to generate rules and a GA to improve simple rules where the initial population is composed of arbitrary rules. Additionally, A. A. Ojugo et al. [16] decided to optimize IDSs by creating rules via a GA. They concluded that the rules created by the GA depend on the selection of fitness function weight values. Furthermore, from all the created rules, they estimated that 80% of them were suitable for detection, and with this result, they concluded that it is better to have a group of good rules than one optimum rule.

## 3      Hybrid Intrusion Detection System

This chapter covers the contributions of this work, the architectural design of the proposed solution, resorting to Unified Modeling Language (UML) [17], 4+1 Architectural View [18], and C4 [19] Models formalisms, a description of the system's operations and a pipeline description of the diagnosis and rule generation.

### 3.1    Contributions

This work contributes to research providing an IDS that stands out for its ML support on populating the knowledge base, which is an operation both intellectually and timely expensive for a human. Not only, but also on its interpretability and explainability, since it justifies the suggested rules, and the diagnosis performed to each asset, and for its capability to act upon arrival of data. Furthermore, since the ML works with empirical knowledge, it can be fine-tuned to the domain where it is deployed. Hence, it can suggest rules that an expert would never think about. The goal of combining these two approaches is to be able to have a SOC engineer filling the knowledge base, either with custom or ML suggested rules, whilst retaining an explainable and interpretable environment.

### 3.2   Architecture of the System

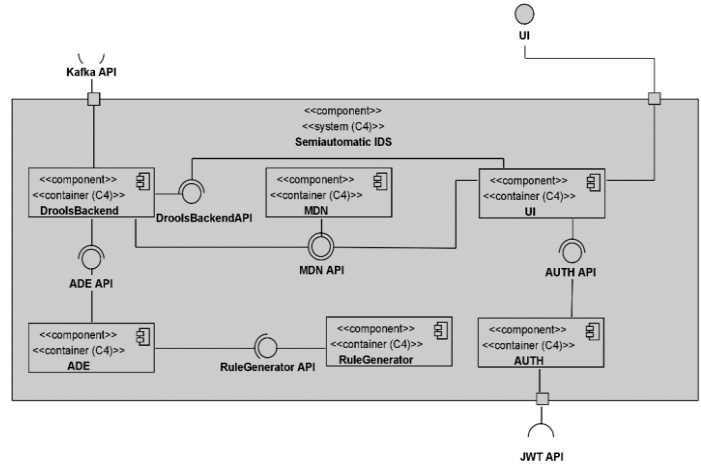

**Fig. 1.** System Architecture

The system follows a microservices-based architecture, as shown in Fig. 1, as it presents many advantages, making the system highly scalable, providing fault isolation, automated deployments, maintainability, reusability, availability and allowing the application to be decentralized [20]. Furthermore, according to the GRASP [21] and SOLID [22] principles, the system's architecture is compliant with the high cohesion, low coupling and the single responsibility principles, since these services hold only limited and focused responsibilities making them small, concise and replaceable.

The solution presented can be broken down into six distinct containers, User Interface (UI); Master Data Network (MDN); Drools Backend; Anomaly Detector & Explainer (ADE); Rule Generator (RG); and Authorization & Authentication (Auth).

The UI is responsible for displaying content and handling user inputs, communicating with the system's backend to trigger functionalities. The MDN container is responsible for all create, read, update and delete (CRUD) operations on the network assets. The Drools Backend is responsible for all CRUD operations on the rules, for diagnosing assets resorting to the Drools engine [23] and justifying it. This container introduces the event-driven paradigm [24], as it uses Apache Kafka [25] to process data in real-time. This service is a subscriber to a Kafka topic, meaning that it receives corresponding data stored in the Kafka server, which is published by a publisher node. In this context, the server works as an event bus since it stores events until they are consumed. This allows the system to act upon the arrival of data. The ADE container is responsible for analyzing the evidence generated to predict the occurrence of an attack, providing a detailed explanation based on the knowledge encoded into the rule-base. The RG is responsible for generating new rules by using the explanations provided by the ADE. Finally, the Auth container is responsible for handling user data, authentication and authorization. The user can interact with all RESTful application programming interfaces (APIs) [26] available in the backend by accessing the UI container.

The UI and Auth containers have been developed in typescript, the first one using React Framework. The MDN, Drools Backend, and Rule Generator have been developed in Java, and the ADE container has been developed in Python. The ML model implemented was the DT with scikit-learn [27]. According to the studies made in [10] and [12], this is a naturally transparent model and, therefore, provides a reasonable level of explainability. The system regards two separate databases, one for storing user-related information and one for data related to the IDS, such as rules, assets, and justifications. This separation was made to increase the security of sensitive data.

### 3.3 System Operations

This section presents the entire operation of this system regarding the interactions between the user and the IDS and its internal operation.

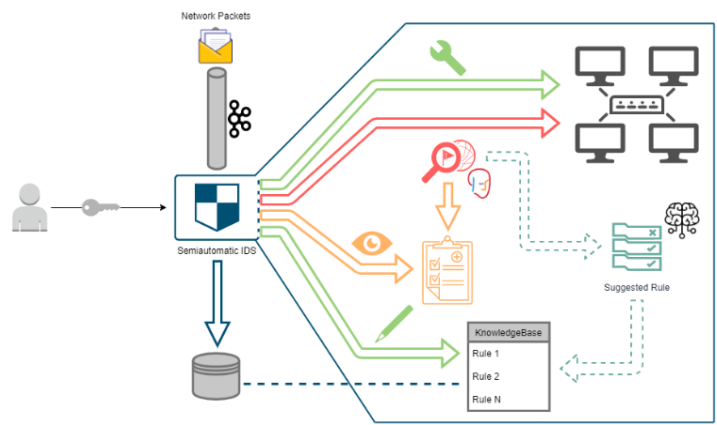

**Fig. 2.** System Operations

As shown in Fig. 2, there are multiple operations that an authenticated and authorized user may perform, such as building a network topology, writing knowledge in the form of rules to the knowledge base, interacting with these, and consulting the assets diagnosis history, as displayed in the green and orange pipelines, respectively. Furthermore, this system can receive network traffic data through a Kafka topic used to diagnose assets, described in the figure as a red pipeline. Each diagnostic leads to a justification and may result in newly suggested rules, represented as a dashed pipeline. The suggested rules may then be accepted. However, it requires user interaction, as the SOC engineer decides about the quality and necessity of said rule to the system's knowledge base.

### 3.4    Pipeline Description

This section presents the steps to diagnose the assets and the generation of rules, shown in Fig. 2 as the red and dashed pipeline, respectively.

**Diagnosis Pipeline –** The **Kafka Producer** starts by sending the packets data to the **Kafka Server**, which the Kafka Consumer then consumes. The data is processed, triggering the calculation of the dynamic evidences. Each dynamic evidence represents information about the sequence of packets that was received. Then these are sent to the Drools Rules Engine. By using this inference engine, it is possible to reach hypotheses and conclusions. The verification of the engine ends when a conclusion is reached or all rules were tested.

When the verification step ends, the matched rules tree is obtained. With these, it is possible to generate a justification and change the status of the assets accordingly. The result is interpreted as follows:

- **If** a conclusion was reached, **then** it is in a danger state;
- **If** no conclusion was reached, but at least one hypothesis was formulated, **then** it is in a warning state;
- **If** none of the predictions were reached, **then** the asset is in a normal state.

**Rules Generation Pipeline –** The dynamic evidences are sent to the ADE container during the diagnostic, which triggers the rule generation pipeline. Upon arrival at the ADE container, the data is processed and then sent to a DT model, making a prediction. In case the prediction is negative, then no rule is created. Otherwise, taking advantage of the DT's explainability, backtracking of the tree is made to understand which conditions lead to it being considered an anomaly. With this traverse of the tree, multiple conditions are matched. These are sent to the RG container that generates the new rule. If the new rule does not exist in the rules database, it is persisted and suggested to the SOC engineer. The current implementation generates as many rules as the number of attack class leaf nodes, each composed of the conditions present in the attack branch.

## 4    Case Study

In cybersecurity, there is a wide variety of cyberattacks. As such, for this case study, a simple *.csv* file was used, containing some test network data that aims to replicate a DDoS attack, particularly an HTTP Flood. The data comprises five features: destination IP; destination port; protocol; packet size; and timestamp.

Two evidence calculators were also implemented, one for the number of packets with HTTP protocol and the other for the number of packets with packet size greater than 4000.0075. These calculators calculate the respective dynamic evidence. Furthermore, the considered IPs were 192.168.1.1, 192.168.1.2, 192.168.1.5 and 192.168.1.40. This last one was targeted to suffer an HTTP Flood attack, according to the sample data in the *.csv* file. Initially, the rules database was empty, so a bootstrap was made to load the network assets and rules described in Table 1.

**Table 1.** Bootstrapped Rules Database

| Rule | Description |
| --- | --- |
| **r50** | **If** the number of packets with HTTP Protocol is higher or equal to 8, **then** there is an **hypothesis** that it may be a DDoS Attack of type HTTP Flood. |
| **r47** | **If** the number of packets with packet size greater than 4000.0075 is higher than 6, **then** there is an **hypothesis** that it may be a DDoS Attack of type Ping of Death. |
| **r3** | **If** the number of packets with packet size greater than 4000.0075 higher than 6 **and** an **hypothesis** that it may be a DDoS Attack of type HTTP Flood has been reached, **then** it can be **concluded** that the attack is indeed a DDoS Attack of type HTTP Flood, and therefore the asset is in **danger**. |

To evaluate the solution, three main validity goals defined were:

- The system works properly when it comes to diagnosing assets;
- The system automatically generates well-formed and coherent rules based on the calculated evidence, resorting to the DT;
- The system is explainable and interpretable for the average user to understand.

### 4.1 System Evidence

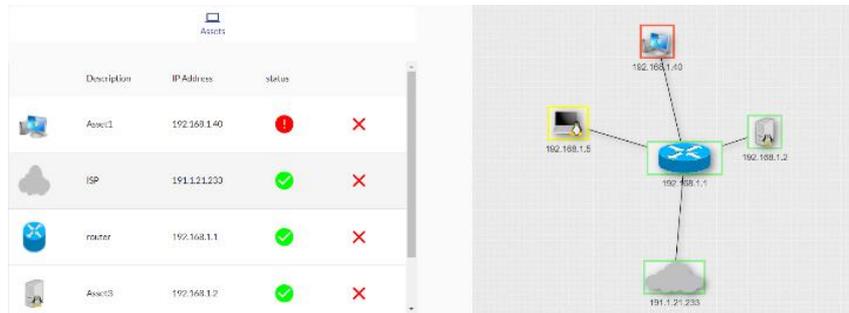

**Fig. 3.** Network Dashboard

While the system was receiving data, it was possible to verify from the network dashboard in Fig. 3, that the status of each asset was changing dynamically and, therefore, that the evidences were being correctly calculated, according to the sample data.

A Hybrid Approach for an Instrusion Detection System      9

**Justification**

Number of packets with packet size greater than 4000.0075 higher than 6 caught by rule r47 which lead to
hypothesis: DDoS Attack of type Ping Of Death
  Number of packets with HTTP protocol higher or equal to 8 caught by rule r50 which lead to
  hypothesis: DDoS Attack of type HTTP Flood
    Number of packets with packet size greater than 4000.0075 higher than 6 and hypothesis: DDoS Attack of type HTTP Flood caught by rule r3 which lead to
    Danger, HTTP Flood

**Fig. 4.** Justification Example

Simultaneously, the diagnosis process was executing and calculating the evidences for each asset, which led to the creation of justifications, such as the one in Fig. 4. By analyzing the justification related to the asset with IP 192.168.1.40, it can be interpreted that it received more than six packets with a packet size greater than four thousand and seventy-five ten thousandths, and the number of packets with HTTP protocol received was greater or equal to eight. As such, fulfilling the conditions side of the rules **r47** and **r50**, which lead the inference engine to generate two hypotheses, one for DDoS attack of type Ping of Death and the other for HTTP Flood. Lastly, since the conditions of the rule **r3** were fulfilled, the system concluded that the asset was in danger, being in the presence of a DDoS attack of type HTTP Flood.

| Name | | | |
|---|---|---|---|
| r3 | RULE CONDITIONS 👁 | ✏ | ✖ |
| PLACEHOLDER 9384f997-1d12-4c1d-9951-3d1ca46c883b | RULE CONDITIONS 👁 | ✓ | ✖ |
| r50 | RULE CONDITIONS 👁 | ✏ | ✖ |
| r47 | RULE CONDITIONS 👁 | ✏ | ✖ |

**Fig. 5.** Rules Dashboard

In the end, by observing the rules dashboard in Fig. 5, which is where the knowledge base is managed, we obtained a new rule and multiple justifications with a conclusive status for each asset. The rule created was accepted and contained the conditions:

- **If** the number of packets with HTTP protocol is lower or equal to 111.5
- **If** the number of packets with packet size greater than 4000.0075 is higher than 4.5

By interpreting the conditions of the newly suggested rule, the SOC engineer can interpret the ML model's prediction. However, the ML model is unable to deduce a right hand side. Therefore, if accepted, the SOC engineer has to fulfil it with a Hypothesis or Conclusion. According to the data sent, this rule can also be considered valuable since it did not yet exist in the system and may detect attacks with fewer packets but greater packet size, characteristic of a low-rate DDoS attack [28].

The justifications created are also comprehensible, enhancing the work's explainability and interpretability. Moreover, being outputted in a tree-like format provides a more intuitive and straightforward understanding of the diagnosis. This can be verified in Fig. 4 for the asset 192.168.1.40.

## 5      Conclusion and Future Work

The proposed solution is able to make use of ML algorithms to enrich expert-written rules contributing to a more dynamic system in which domain knowledge is constantly being updated. It is also able to detect anomalies resorting to a knowledge-based approach. Furthermore, this system was able to achieve explainability and interpretability through the justifications of each diagnosis; the use of an explainable ML model, which provides justifications to a certain prediction in the form of a suggested rule; and an intuitive UI to display the respective information.

As future work, it could be helpful to expand its knowledge base and experiment with other promising ML models to be employed, so it can become an even more robust IDS and detect more attacks.

**Acknowledgements.** This work was partially supported by the Norte Portugal Regional Operational Programme (NORTE 2020), under the PORTUGAL 2020 Partnership Agreement, through the European Regional Development Fund (ERDF), within project "Cybers SeC IP" (NORTE-01-0145-FEDER-000044). This work has also received funding from the following projects: UIDB/00760/2020, UIDP/00760/2020 and CyberFactory# 1 (Refª: NORTE-01-0247-FEDER-40124).

A Hybrid Approach for an Instrusion Detection System    11